\documentclass[amsmath,amssymb,twocolumn,aps,prd,10pt]{revtex4-2}
\usepackage[T1]{fontenc}
\usepackage{graphicx}
\usepackage{subfig}
\usepackage{epsfig}
\usepackage{dcolumn}
\usepackage{multirow}
\usepackage{rotating}
\usepackage{verbatim}
\usepackage{bm}
\usepackage{color}
\usepackage{soul}
\usepackage{float}
\usepackage{hyperref}

\def\lsim{\raise0.3ex\hbox{$<$\kern-0.75em\raise-1.1ex\hbox{$\sim$}}}
\def\gsim{\raise0.3ex\hbox{$>$\kern-0.75em\raise-1.1ex\hbox{$\sim$}}}

\def\pom{{I\!\!P}}
\def\reg{{I\!\!R}}

\def\beqa{\begin{eqnarray}}
\def\eeqa{\end{eqnarray}}

\begin{document}

\title{Dilepton production through timelike Compton scattering within the $k_T$-factorization approach}
\author{G. M. Peccini} 
\email{guilherme.peccini@ufrgs.br}
\author{L. S. Moriggi}
\email{lucas.moriggi@ufrgs.br}
\author{M. V. T. Machado}
\email{magnus@if.ufrgs.br}
\affiliation{High Energy Physics Phenomenology Group, GFPAE. Institute of Physics, Federal University of Rio Grande do Sul (UFRGS)\\
Caixa Postal 15051, CEP 91501-970, Porto Alegre, RS, Brazil}

\begin{abstract}
In this work we consider the dilepton production via timelike Compton scattering (TCS) in electron-proton and proton-proton collisions. In particular, the differential cross section in terms of the dilepton invariant mass and rapidity is computed within the $k_T$-factorization approach. Besides, we utilize distinct unintegrated gluon distributions (UGD) in order to compare their  impact on the differential cross section of TCS in $pp$ ($ep$) collisions evaluated at the LHC (LHeC), HL-LHC (LHeC), HE-LHC (LHeC) and FCC-hh (eh) center-of-mass energies.
\end{abstract}

\maketitle

\section{Introduction} 

Dilepton production can occur through several mechanisms, being the leading one the ordinary Drell-Yan process. The second most important contribution comes from photon fusion, i.e., $\gamma \gamma \to \ell^+ \ell^-$, which is used for controlling the luminosity at the LHC. In addition, single and double diffractive Drell-Yan also produce dileptons via different interactions, such as Pomeron - Pomeron ($\pom \pom$), Pomeron - Reggeon ($\pom \reg$), Reggeon - Reggeon ($\reg \reg$), Pomeron - proton ($\pom p$) and Reggeon - proton ($\reg p$) reactions \cite{Kubasiak:2011xs,Goncalves:2018gca}. We can still have the reactions $\gamma \pom$ and $\pom \gamma$, where the underlying process is the time-like Compton scattering (TCS). At last, it should be mentioned that the Bethe-Heitler (BH) mechanism contributes at the amplitude level to the physical process of photoproduction of heavy lepton pairs, and it is  known that the BH contribution (and its interference with TCS) is large in contrast to timelike Compton scattering itself. 

Timelike Compton scattering has been commonly investigated within the formalism of Generalized Parton Distributions (GPDs) \cite{Diehl:2003ny,doi:10.1146/annurev.nucl.54.070103.181302,Belitsky:2005qn} (see also, for example, Refs. \cite{Boer:2015fwa,Berger:2001xd,Moutarde:2013yua}). One of the goals in the study of these distributions is  to understand how quarks and gluons assemble themselves to hadrons \cite{Mueller:1998fv,Blumlein:1999sc,Ji:1996ek,Radyushkin:1997ki}. Since the cleanest reactions to obtain the GPDs are the DVCS (deeply virtual Compton scattering) and TCS, studying the latter through distinct reactions could be relevant for their determination. Being the \textquotedblleft inverse \textquotedblright  process of the former, in TCS a quasi-real photon interacts with a proton and the final state after the scattering is an outgoing proton and a timelike virtual photon that subsequently decays into a lepton pair. Namely, the process is the following: $\gamma p \to \gamma^*p$. In the context of GPDs, recently the TCS amplitudes and associated observables have been investigated in leading-twist approximation \cite{Grocholski:2019pqj} and a careful analysis is done in order to substantially reduce the model dependence. The dilepton production from TCS was adressed in ultraperipheral collisions (UPCs) at a fixed-target experiment (AFTER@LHC) using the nucleon and ion beams in Ref. \cite{Lansberg:2015kha}. Moreover, the linearly polarized photon beam has been considered in \cite{Goritschnig:2014eba}, where new observables were proposed and the impact on determination of polarized  GPDs has been studied. Yet, the NLO corrections to the timelike (TCS), spacelike (DVCS) and double deeply virtual Compton scattering (DDVCS) amplitudes have been fully demonstrated in \cite{Pire:2011st}.

The process has been also investigated within the color dipole approach. In Ref. \cite{Machado:2008zv}, the cross section was first computed using a spacelike approximation for $ep$ and $eA$ collisions. It was a straightforward application of a previous work on the diffractive photoproduction of $Z^0$ done in Ref. \cite{Goncalves:2007vi} and it is complementary to the predictions for nuclear DVCS \cite{Machado:2008tp}. The comparison between the inclusive and exclusive dilepton photoproduction was done in Ref. \cite{Mariotto:2013qsa}. The wave function for an outgoing photon with timelike $q^2>0$ was derived in \cite{Motyka:2008ac}. It was shown that the cross section calculation involves a strong oscillatory integrand, which was solved by taking analytic continuation to complex transverse dipole size, $r$,  with a suitable integration contour. This difficulty does not appear if the transverse momentum space is considered. The TCS process is deeply connected with the DDVCS process, $\gamma^*p\rightarrow \gamma^*p\rightarrow (\ell^+ \ell^-)p$, which was treated in the context of dipole framework in \cite{Kopeliovich:2010xm} considering the scattering on nucleons and nuclei.

It is well known that, at asymptotically high energies, BFKL (Balitsky-Fadin-Kuraev-Lipatov) \cite{Balitsky:1978ic,Kuraev:1977fs} evolution describes the gluon dynamics. The corresponding evolution equation describes the $x$ behavior of the unintegrated distribution. The results coming from DGLAP (Dokshitzer-Gribov-Lipatov-Altarelli-Parisi) \cite{Dokshitzer:1977sg,Gribov:1972ri,Altarelli:1977zs} evolution coincide with those from BFKL in the double logarithmic limit. Both approaches predict strong rise on $F_2$ at small Bjorken variable $x$, as measured at HERA. However, BFKL evolution predicts a strong power-like rise. Similarly to the collinear factorization, one can factorize an observable into a convolution of process-dependent hard matrix elements with universal parton distributions, but here the virtuality and transverse momentum of the propagating gluon are no longer ordered and then the matrix elements need to be taken off-shell, which leads to the fact that the  convolution also occurs over transverse momentum through unintegrated parton distributions (uPDF). This formalism is the so-called $k_T$-factorization \cite{Andersson:2002cf}. Hereafter, in most cases we will refer to uPDFs simply as UGDs (unintegrated gluon distributions) since at the small-$x$ regime the dominant partons are the gluons. 

At small-$x$ (high energies), since $\Delta y \sim ln (1/x)$, the evolution of parton distributions go along over a large region of rapidity. In this context, the effects of finite transverse momenta of partons can become progressively important. Thus, the cross sections may be factorized into a $k_T$ partonic cross section and an unintegrated parton distribution, $\phi (z, k_T^2)$. For inclusive processes, it is calculated as \cite{Andersson:2002cf}

\begin{equation}
\sigma = \int \frac{dz}{z}    d^2k_T \  \hat{\sigma}(\frac{x}{z},k_T^2)\phi (z,k_T^2)
\end{equation}

In this work, we focus on TCS and corresponding dilepton production in  both $ep$ and $pp$ reactions within the  $k_T$-factorization approach. In Ref. \cite{Schafer:2010ud}, the process was first  calculated in such a formalism and  the differential cross sections for dilepton production as a function of invariant mass and energy have been analysed for electron-proton scattering process. There, the authors utilized an unintegrated gluon distribution proposed in Ref. \cite{Ivanov:2000cm}. 

There are not many studies regarding TCS in literature so far. Therefore, this subject has not been substantially explored yet and, in that sense, further investigations could be highly relevant. The aim here is to extend the analysis carried out in Ref. \cite{Schafer:2010ud} by taking into account other UGDs in order to single out the model dependence and perform predictions for the future  high energy $ep$ colliders LHeC, HL-LHeC, HE-LHeC and FCC-eh. The main goal is to compare the differential cross sections (from different UGDs) with respect to the dilepton invariant mass and rapidity distributions, as well as the total production cross section. Furthermore, we also do the investigation for proton-proton collisions, in which one of the mechanisms for dilepton production is the photon-Pomeron interaction where TCS is present. 

This paper is organized as follows. In the next section, we summarize the derivation of the cross section for dilepton production by TCS in the transverse momentum space for electron-proton and proton-proton collisions. In Section III, we introduce the different UGDs that will be utilized in this work and apply them to the  calculations, discussing the model dependence in the expressions. This aims to understand the theoretical uncertainties and propose observables to be measured at future high energy and high luminosity $ep$ and $pp$ machines.  Finally, in Section V we outline the paper and expose our main conclusions.

\section{Dilepton production via timelike Compton scattering (TCS)}

\subsection{TCS in $ep$ collisions}

Initially, we address the exclusive dilepton production through TCS in electron-proton collisions ($\gamma p \to \gamma^* p$). We will adopt the formalism proposed in Ref. \cite{Schafer:2010ud}, where the imaginary part of the TCS amplitude is calculated in terms of the  unintegrated gluon function within the $k_T$-factorization approach. The underlying process is the color dipole $q\bar{q}$ interaction with the proton producing an exclusive final state where a QCD Pomeron is exchanged in $t$-channel. The amplitude is given below:
\begin{equation}
ImM_f(W^2,\kappa^2,z) = \int _0 ^ \infty \frac{d^2 k_{\perp}}{k_{\perp}^2} \phi(x,k_{\perp}^2) \alpha_s(\mu ^2)
\end{equation}
\begin{equation}
\times \big [C_{0f}(z,\kappa^2)D_{0f}(\kappa^2,k_{\perp}^2)+C_{1f}(z,\kappa^2)D_{1f}(\kappa^2,k_{\perp}^2)\big ] \ ,
\label{eq:main}
\end{equation}
where $k_{\perp}^2$ stands for the transverse momentum squared of the gluons, while $\kappa^2$ represents the transverse momentum squared of the quarks. In the expression above, $z$ is the longitudinal momentum fraction carried by the quarks and $\phi(x,k_{\perp}^2)$ is the UGD. The running coupling, $\alpha_s(\mu ^2)$, is being taken at $\mu ^2 =  \mathrm{max} (\kappa ^2 +m_f^2,k_{\perp}^2)$. Adopting the prescription of Ref. \cite{Ivanov:2000cm}, if $\alpha_s$ exceeds 0.82, it is frozen at this value in order to assure perturbative calculation. The functions $C_{0f}$, $D_{0f}$, $C_{1f}$ and $D_{1f}$ are defined as follows:
\begin{eqnarray}
C_{0f}(z,\kappa^2)&=&m_f^2,\quad D_{0f}(\kappa^2,k_{\perp}^2)=\frac{1}{\alpha}-\frac{1}{\beta}, \\
C_{1f}(z,\kappa^2)&=&[z^2+(1-z)^2]\frac{\kappa^2}{\alpha}, \\
D_{1f}(\kappa^2,k_{\perp}^2)&=&1- \frac{\alpha}{2\kappa^2}\bigg (\frac{\kappa^2-m_f^2-k_{\perp}^2}{\beta}+1\bigg ) \ ,
\end{eqnarray}
where $m_f$ is the quark mass of flavor $f$ and $\alpha$ and $\beta$ are given by
\begin{eqnarray}
\alpha=m_f^2+\kappa^2, \quad    \beta=\sqrt{(\kappa^2-m_f^2-k_{\perp}^2)^2+4m_f^2\kappa^2}.  
\end{eqnarray}

Having $Im\,M_f$, we shall define the spectral distribution, which is regarded to the diffractive amplitude for the $\gamma p \to q \bar{q} p$ (the virtual photon is being taken as a quark-antiquark pair) transition as

\begin{eqnarray}
Im\,\mathcal{M}_f(W^2,M_{q\bar{q}}^2)&=&\frac{1}{\pi M_{q\bar{q}}^2}\int _0 ^{\kappa_{\mathrm{max}}^2}\frac{d^2\kappa}{\sqrt{1-4\left(\frac{\alpha}{M_{q\bar{q}}^2}\right)}} \nonumber \\
&\times & Im\,M_f(W^2,\kappa^2,z)  \ , 
\label{eq:spectral}
\end{eqnarray}
where $\kappa_{\mathrm{max}}^2 = (0.25M_{q\bar{q}}^2-m_f^2)$.

In Eq. (\ref{eq:spectral}), $W^2$ and $M_{q\bar{q}}^2$ represent the photon-proton center-of-mass energy squared and the dipole invariant mass squared, respectively. The latter is related to $\kappa^2$ and $z$ by $M_{q\bar{q}}^2=(\kappa^2+m_f^2)/z(1-z)$. The UGD is taken at $x=\delta \frac{M_{q\bar{q}}^2}{W^2}$ in order to correct the skewedness effects, where $\delta=0.41$. Here we are following closely Ref. \cite{Schafer:2010ud}, but different prescriptions for the skewedness corrections could be employed. For instance, a skewedness factor $R_g$ using the Shuvaev et al. expression for gluons \cite{Shuvaev:1999ce} can be multiplied to the amplitude or the prescription of Harland-Lang \cite{Harland-Lang:2013xba} can be considered, where the skewed gluon density is simply related to the gluon GPD. The off-diagonal correction is one of the theoretical uncertainties in the calculations.

The TCS scattering amplitude is computed as
\begin{eqnarray}
 & & \mathcal{A}_f^{TCS} (\gamma p \to \gamma^{*}(M_{\ell^+\ell^-}^2) p ) = \frac{4}{\pi} \alpha_{em} e_f^2 \nonumber \\
&\times & \int _{4m_f^2}^{\infty} \frac{\mathcal{M}_f(W^2,M_{q\bar{q}}^2)}{M_{q\bar{q}}^2-M_{\ell^+\ell^-}^2-i\epsilon}  \ dM_{q\bar{q}}^2.
\label{eq:imag_amp}
\end{eqnarray}
The integration in the domains of $M_{\ell^+\ell^-}^2 > 4m_f^2$ and $M_{\ell^+\ell^-}^2 < 4m_f^2$  is written in the following way:
\begin{equation*}
Im\mathcal{A}_f^{TCS}= \frac{4  \alpha_{em}e_f^2}{\pi} \Big [ \Theta (M_{\ell^+\ell^-}^2 - 4m_f^2)
\end{equation*}
\begin{equation*}
\times \big (PV \int _{4m_f^2}^{\infty}\xi (W^2, M_{q\bar{q}}^2,M_{\ell^+\ell^-}^2)  \ dM_{q\bar{q}}^2
\end{equation*}
\begin{equation*}
+\pi Re\mathcal{M}_f(W^2,M_{\ell^+\ell^-}^2)\big )+ \Theta (4m_f^2 - M_{\ell^+\ell^-}^2)
\end{equation*}
\begin{equation}
\times \int _{4m_f^2}^{\infty} \xi (W^2,M_{q\bar{q}}^2,M_{\ell^+\ell^-}^2)  \ dM_{q\bar{q}}^2\Big ] \ ,  
\label{eq:theta}     
\end{equation}
where  PV stands for the Cauchy Principal Value and the auxiliary function $\xi$ has been defined as
\begin{equation}
\xi (W^2, M_{q\bar{q}}^2,M_{\ell^+\ell^-}^2) =  \frac{Im\mathcal{M}_f (W^2,M_{q\bar{q}}^2)}{M_{q\bar{q}}^2 - M_{\ell^+\ell^-}^2} \ .
\end{equation}

At this point, some considerations are in order. The upper bound of the integral in Eq. (\ref{eq:imag_amp}) leads to contributions $x \simeq 1$ or so (see the definition of $x$). These contributions are suppressed by the $1/M_{q\bar{q}}^2$ factor in the spectral distribution, Eq. (\ref{eq:spectral}), and by the large $x$ threshold factor $(1-x)^n$ present in the phenomenological UGDs. 

Regarding Eq. (\ref{eq:theta}), $\Theta$ denotes the Heaviside function and $e_f^2$ is the squared quark charge of flavor $f$. Analogously, the real part of the amplitude is evaluated by 

\begin{equation*}
Re\mathcal{A}_f^{TCS}= \frac{4  \alpha_{em}e_f^2}{\pi} \Big [ \Theta (M_{\ell^+\ell^-}^2 - 4m_f^2)
\end{equation*}
\begin{equation*}
\times \big (PV \int _{4m_f^2}^{\infty}\eta (W^2,M_{q\bar{q}}^2,M_{\ell^+\ell^-}^2)  \ dM_{q\bar{q}}^2
\end{equation*}
\begin{equation*}
-\pi Im\,\mathcal{M}_f(W^2,M_{\ell^+\ell^-}^2)\big )
+ \Theta (4m_f^2 - M_{\ell^+\ell^-}^2) 
\end{equation*}
\begin{equation}
\times \int _{4m_f^2}^{\infty} \eta (W^2,M_{q\bar{q}}^2,M_{\ell^+\ell^-}^2)  \ dM_{q\bar{q}}^2\Big ]       \ ,
\end{equation}
In the previous expression, the definition of $\eta (W^2,M_{q\bar{q}}^2,M_{\ell^+\ell^-}^2)$  is the following:
\begin{equation}
\eta (W^2,M_{q\bar{q}}^2,M_{\ell^+\ell^-}^2) =  \frac{Re\mathcal{M}_f (W^2,M_{q\bar{q}}^2)}{M_{q\bar{q}}^2 - M_{\ell^+\ell^-}^2} \ . 
\end{equation}
The function $ReM_f$ is obtained by using the dispersion relation,
$\rho = Re\,M_f/ImM_f$. The $\rho$ parameter is given by $\rho = \tan \left( \frac{\pi}{2} \lambda_{\mathrm{eff}}\right)$, where $\lambda_{\mathrm{eff}}=\partial \ln(Im M_f)/\partial \ln(W^2)$.
Taking into consideration the definition of the variable $x$ alongside Eq. (\ref{eq:main}), the derivative in the above expression is easily obtained. For simplicity, the diffraction cone approximation will be used, which enables one to embed a $t$ dependence by means of the following factorization:
\begin{equation}
A_f^{TCS}(W,t)=A_f^{TCS}(W) e^{Bt} \ ,
\end{equation}
where $B$ is the slope parameter. In this work, we will adopt $B=4 \ GeV^{-2}$. By utilizing the optical theorem, evaluating the total cross section for the $\gamma p \to \gamma^{*}p $ process is straightforward,
\begin{eqnarray}
\sigma^{TCS}(\gamma p \to \gamma ^* p ) = \frac{[Im (A^{TCS})]^2 + [Re (A^{TCS})]^2}{16 \pi B} \ ,
\end{eqnarray}
where $Im A^{TCS} = \sum Im A^{TCS}_f$ and $Re A^{TCS}=\sum Re A^{TCS}_f $.
Finally, we can express the differential cross section in terms of the dilepton invariant mass distribution, i.e, 

\begin{equation}
\frac{d \sigma (\gamma p \to \ell^+\ell^- p)}{dM_{\ell^+\ell^-}^2} = \frac{\alpha_{em}}{3 \pi M_{\ell^+\ell^-}^2} \sigma^{TCS}(\gamma p \to \gamma ^* p ) \ .
\label{eq:difXsec}
\end{equation}
Here, we will only consider the charm and the light quarks, whose values are taken according to the corresponding UGD model applied.
Lastly, one may integrate Eq. (\ref{eq:difXsec}) in order to get the cross section integrated over the dilepton invariant mass,  $M^2_{\ell^+\ell^-}$, i.e.,

\begin{equation}
\sigma_{tot}(\gamma p \to \ell^+\ell^- p)=\int_{{(M^2_{\ell^+\ell^-})}_{\mathrm{min}}}     ^{\infty} \frac{d\sigma}{dM_{\ell^+\ell^-}^2} dM_{\ell^+\ell^-}^2   \ ,
\end{equation} 
where ${(M^2_{\ell^+\ell^-})}_{\mathrm{min}}$ is the (cut) minimum invariant mass of the lepton pair.

\subsection{TCS in $pp$ collisions}

In case of $pp$ collisions, the production of lepton pairs via TCS is carried out by photon-Pomeron ($\gamma \pom$) and Pomeron-photon ($\pom \gamma$) mechanisms. Following Ref. \cite{Kubasiak:2011xs}, the $\gamma \pom + \pom \gamma$ contribution for the amplitude of the $pp \to pp \ \ell^+\ell^-$ process may be obtained within the equivalent photon approximation (EPA) and reads as
\begin{eqnarray}
\frac{d\sigma }{d M^2_{\ell^+
\ell^-} dy_{\mathrm{pair}}} & = & k_{+}   \frac{dn (k_+)}{dk_+} \ \frac{d \sigma ^{TCS}}{dM^2_{\ell^+\ell^-}} (W_+) \nonumber \\
& + & k_-     \frac{d n(k_-)}{dk_-} \frac{d\sigma^{TCS}}{dM^2_{\ell^+\ell^-}} (W_-) \ ,
\end{eqnarray}
where $k$ is the photon energy, $dn (k)/dk$ is the photon flux and $y_{pair}$ is the dilepton rapidity. The subscripts $+$ and $-$ are related to the $\gamma \pom$ and $\pom \gamma$ subprocesses, respectively. The flux expression will be extracted from Ref. \cite{Drees:1988pp}, in which it is written as
\begin{equation*}
\frac{d n(k)}{dk}=\frac{\alpha_{em}}{2 \pi } \bigg [1+\bigg (1-\frac{2k}{\sqrt{s}} \bigg )^2\bigg ]
\end{equation*}
\begin{equation}
\times \bigg ( ln \ \chi  -\frac{11}{6}+\frac{3}{\chi}-\frac{3}{2 \chi^2}+\frac{1}{3\chi^3} \bigg )  \ .
\end{equation}
 The quantity $\sqrt{s}$ is the center-of-mass energy of the $pp$ system and the parameter $\chi$ is defined as $\chi=1+(Q_0^2/Q_{min}^2)$ with $Q_0^2 =0.71 \ GeV^2$ and $Q_{min}^2=k^2 / \gamma_L^2$, where $\gamma _L=\sqrt{s}/2m_p$.  
 
Given the definitions of rapidity and $s$ (Mandelstam variable), one can express the following relations:
\begin{eqnarray}
k_{\pm}=\frac{M_{\ell^+\ell^-}}{2} \ e^{\pm y_{\mathrm{pair}}}\ \quad W^2_{\pm}= 2 k_{\pm} \sqrt{s} \ .
\end{eqnarray}
The expression above relates the photon-proton center-of-mass energy to the proton-proton one.

From the experimental point of view, the production of exclusive  dilepton events is relatively understood. For instance, ATLAS collaboration has recently performed measurement at 13 TeV for a dimuon invariant mass of $12 < M_{\ell^+\ell^-}< 70$ GeV \cite{Aaboud:2017oiq}. Also, CMS collaboration \cite{Cms:2018het} has measured proton-tagged events at the same energy for exclusive dilepton produced at
midrapidity with $M_{\ell^+\ell^-}> 110$ GeV and one of the two scattered protons is measured in precision proton spectrometer (CT-PPS). ATLAS has reported similar measurement of forward proton scattering in association with dileptons produced via $\gamma \gamma$ fusion with a significance higher than $5\sigma$ \cite{Aad:2020glb}. On the phenomenological side, in the new SuperChic 4 Monte Carlo \cite{Harland-Lang:2020veo} photon-initiated production in proton-proton collisions has been implemented. The code takes into account the different contributing channels, including proton dissociation.

\section{Results and Discussion}

\begin{figure*}[t]
\centering
    \includegraphics[width=0.8\textwidth]{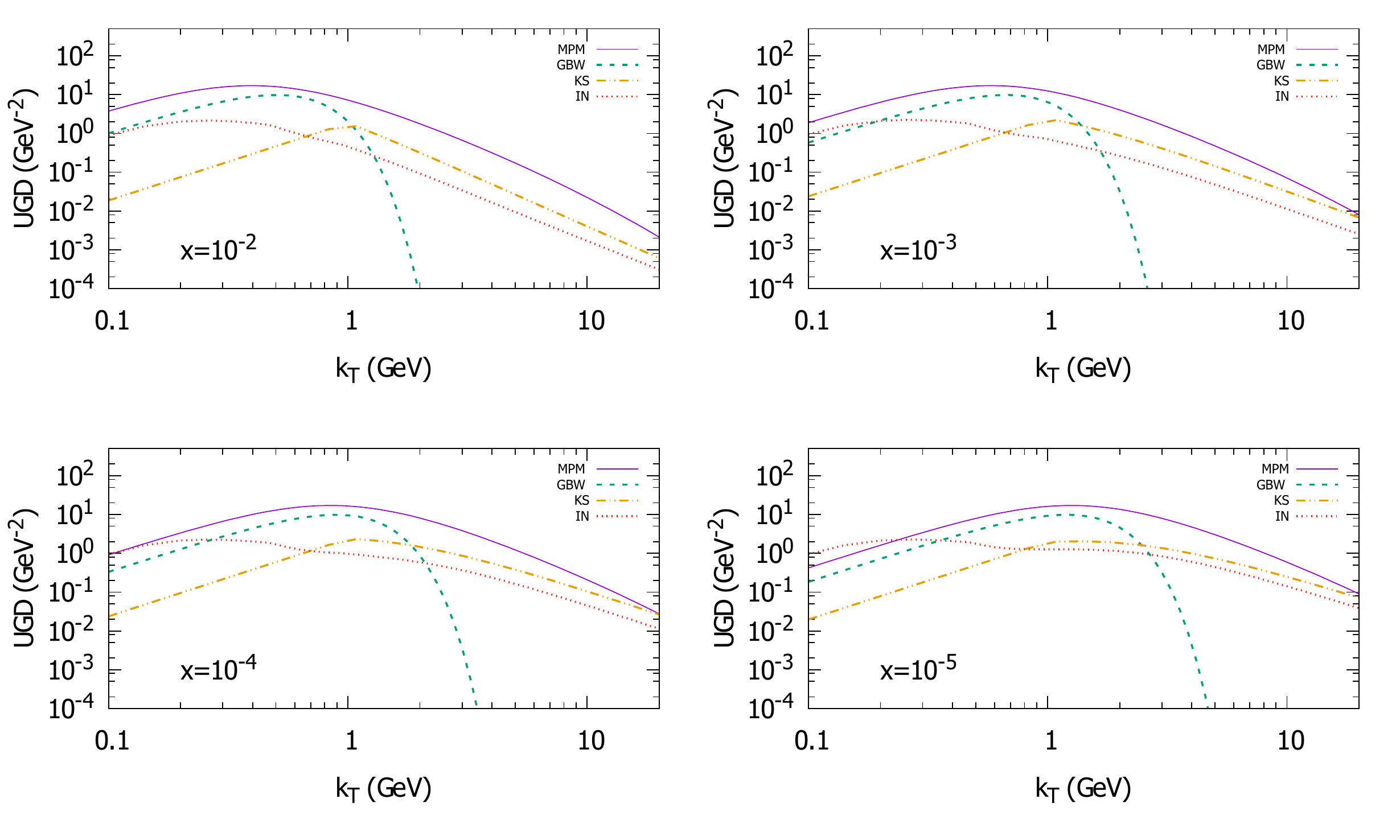}
    \caption{Comparison between the considered UGDs (MPM, GBW, KS and IN). The unintegrated gluon distribution is shown as a function of gluon transverse momentum for fixed values of Bjorken variable.}
    \label{fig:ugd}
\end{figure*}

\begin{figure*}[t]
\centering
    \includegraphics[width=0.8\textwidth]{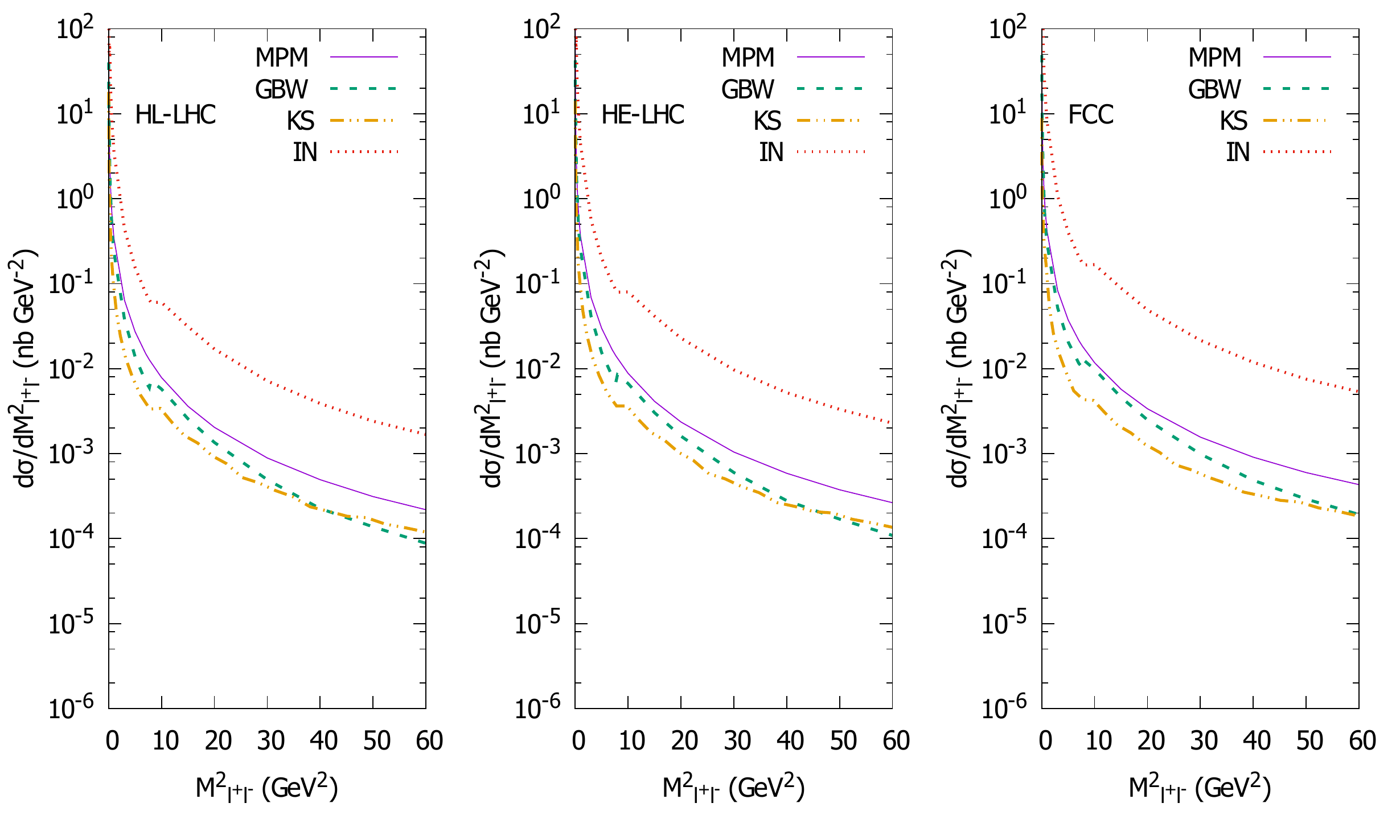}
    \caption{Differential cross section of TCS as a function of dilepton invariant mass, $M_{\ell^+\ell^-}$, for $ep$ collisions at the LHeC/HL-LHeC, HE-LHeC and FCC-eh energies.}
    \label{fig:dsdmep}
\end{figure*}

As pointed out in the Introduction, we aim to calculate the TCS process in $ep$ and $pp$ collisions within the $k_T$-factorization approach. To do so, an unintegrated gluon distribution is needed as the non-perturbative input of the formalism. In that sense, in this work we will consider four UGDs containing different physical informations.  Initially, we take the KS (Kutak-Sapeta) UGD, specifically its non-linear set \cite{Kutak:2012rf}, which takes into account parton saturation effects. The KS distribution was imported from the TMDlib (Transverse Momentum Dependent parton distributions library) \cite{Hautmann:2014kza}, which provides a large number of uPDFs.

Due to non-positive definite kernel, the basic formulation of the NLO BFKL equation is unstable. In order to stabilize it, one should resume a subset of higher order corrections. In Ref. \cite{Kwiecinski:1997ee}, the authors took the higher order corrections from consistency constraint on emission of real gluons. In addition, more corrections are performed by running the constant coupling and other contributions come from non-singular pieces of the DGLAP splitting functions. In this context, the authors of Ref. \cite{Kutak:2012rf} proposed the KS model, whose expression may be seen in the quoted reference.

The next UGD considered has been developed in Ref. \cite{Moriggi:2020zbv} (we will name it as MPM hereafter), where the authors utilized the geometric scaling framework to build an  expression for the gluon unintegrated function that depends on the variable $\tau$, being $\tau=\frac{k_T^2}{Q_s^2}$, where $k_T^2$ is the transverse momentum of the gluons and $Q_s^2$ is the saturation scale. Along with the variable $\tau$, the MPM has also three other parameters (see the quoted reference for details). Therein, in order to avoid the divergence of jet production in the infrared sector (IR), the saturation scale is taken as an  effective regulator of the gluon propagator compatible with a Yukawa potential, $\phi (k_T^2) \sim \alpha_s k_T^2 / (1+k_T^2/\mu^2)$, leading to a distribution of the following form:
\begin{equation}
\phi_{\mathrm{MPM}} (x,k_T^2)= \frac{3 \sigma_0}{4 \pi ^2 \alpha_s}    \frac{(1+\delta n)}{Q_s^2}\frac{k_T^2}{ (1+ \frac{k_T^2}{Q_s^2})^{2+\delta n}},
\end{equation}
where $Q_s^2=(x_0/x)^{0.33}$ and $\delta n = a \tau ^b$. The parameters $\sigma_0$, $x_0$, $a$ and $b$ were fitted against DIS data for $x<0.01$. The model describes simultaneously the DIS data at small-$x$ and the spectra of produced hadrons in $pp/p\bar{p}$ collisions. The MPM model is based on geometric scaling arguments and Tsallis-like behavior of the measured spectra. Furthermore, it has no dependence on the scale $\mu ^2$ and a coupling constant, $\alpha_s=0.2$, is assumed.

\begin{table}[t]
\begin{center}
\begin{tabular}{|l|c|c|c|}
\hline 
Collider & $E_e$ (GeV) & $E_p$ (TeV) & $\sqrt{s}$ (TeV) \\
\hline
LHeC/HL-LHeC    & 60 & 7 &  1.3 \\
HE-LHeC & 60 & 13.5 &  1.7 \\
FCC-eh     & 60 & 50 &  3.5 \\
\hline
\end{tabular}
\end{center}

\caption{Estimated energies of the beams at future electron-proton colliders (LHeC/HL-LHeC, HE-LHeC and FCC-eh).}
\label{tab:1}
\end{table}

The third UGD is based on the GBW (Golec-Biernat-W\"{u}sthoff) parametrization \cite{GolecBiernat:1998js}. Having parameters fitted from DIS data at small-$x$, the expression is analytical and given by \cite{GolecBiernat:1998js}:
\begin{equation}
\phi_{GBW}(x,k_{T}^2)=\frac{3\sigma_0}{4 \pi^2 \alpha_s }\bigg ( \frac{k_{T}^2}{Q_s^2}\bigg )e^{-\frac{k_{T}^2}{Q_s^2}}.
\end{equation}
Above, the variable $Q_s$ is the saturation scale and its value is $Q_s^2=(x_0/x)^{\lambda}$, while $\sigma_0=27.32 \ mb$, $\lambda=0.248$ and $x_0=4.2 \times 10^{-5}$ \cite{Golec-Biernat:2017lfv}. The model above holds the small-$x$ region and then presents the geometric scaling property with dependence on the ratio $k_{T}^2/Q_s^2$. Likewise the MPM model, the GBW parametrization has no dependence on $\mu ^2$ and $\alpha_s=0.2$.

\begin{figure*}[t]
\centering
    \includegraphics[width=0.8\textwidth]{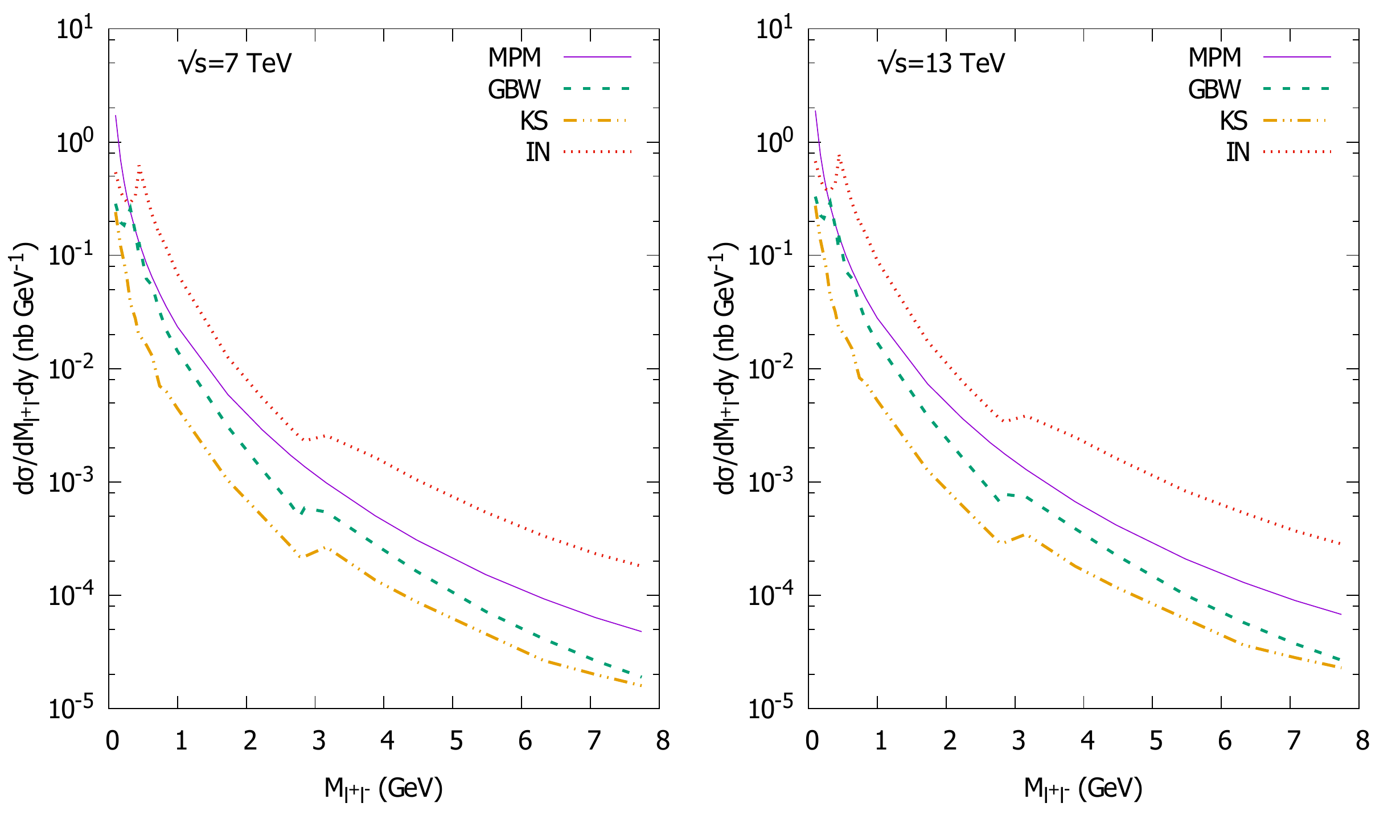}
    \caption{Differential cross section of TCS in terms of dilepton invariant mass distribution for $pp$ collisions at the LHC energies of 7 TeV (left panel) and 13 TeV (right panel) at central rapidity (y=0).}
    \label{fig:dsdmpp}
\end{figure*}

Finally, we also accounted for the UGD proposed in Ref. \cite{Ivanov:2000cm} (we will call it IN hereafter), which is the distribution used in \cite{Schafer:2010ud} . This parametrization has been separated into two parts, namely the soft and hard ones. The latter is also divided in two underlying contributions. For large $Q^2$, the UGD is simply taken as the derivative of the gluon PDF (for instance, GRV, CTEQ, MRS...) with respect to $ln \ Q^2$. On the other hand, for $\kappa^2 \lesssim 1/R_c$   ($\kappa^2$ is the gluon transverse momentum), the UGD dependence on $\kappa^2$ is similar to the Yukawa screened flux of photons in the positron (see Eq. (4) of Ref. \cite{Ivanov:2000cm}). The quantity $Q_c$ is inversely proportional to  the screening radius, $Q_c \sim 1/R_c$, with $R_c \sim 0.27 \ fm$. This variable denotes the propagation/screening of the perturbative color fields (Yukawa-Debye screening). Regarding the soft part of the ansatz, it considers large dipoles in which the dipole cross section does not depend on the energy. In order to verify the explicit expression of this UGD and the corresponding detailed analysis, see Ref. \cite{Ivanov:2000cm} and Eq. (43) therein. The fact is that the IN distribution is not based on saturation physics arguments and the transition between the hard and soft regimes is set by the fixed IR scale, $Q_c$.  The model is quite successful and describes correctly the structure functions $F_2, \,F_L, \,F_2^{c\bar{c}}$ and gives reasonable predictions for exclusive vector meson production \cite{Cisek:2014ala,Ivanov:2004ax,Caporale:2005rj}, as well as the exclusive jet production \cite{Szczurek:2000pj}.

A comparison between the considered UGDs -  MPM (solid lines), GBW (dashed lines), KS (dot-dashed lines) and IN (dotted lines) -  is done in Fig. \ref{fig:ugd}, where they are presented  as a function of gluon transverse momentum, $k_T$, for fixed values of Bjorken variable: $x = 10^{-2}, \,10^{-3}, \,10^{-4}, \,10^{-5}$. It is noticed that the behavior in $k_T> 1$ GeV is somewhat similar for the MPM, KS and IN models despite their distinct overall normalizations. In these UGDs that region is driven by the DGLAP-like behavior for the integrated gluon distribution. The GBW model presents the already known fall-off for large momentum. The transition to the soft region at small $k_T$ is different in each model. In the IN parametrization, it occurs for a fixed momentum value, $k_T = Q_c$, whereas it is dynamical for models based on saturation formalism. The critical line is established by the saturation scale, $Q_s(x)$. We anticipate that the predicted cross section will have significant variability in their overall normalizations.

\begin{table*}[t]
\begin{center}
\begin{tabular}{|l|c|c|c|c|c|c|}
\hline
  & LHeC|HL-LHeC & Event Rate ($\times 10^8$) & HE-LHeC & Event Rate ($\times 10^8$) & FCC-eh & Event Rate ($\times 10^8$)\\
\hline
GBW     & 1.569 (0.413)&0.130|1.042  & 1.664 (0.452)&1.711  &  1.991 (0.574)& 2.716  \\
MPM    & 1.959 (0.641)	& 0.202|1.617 & 2.070 (0.693)	 &2.623 &2.378 (0.842)&3.983    \\
KS     & 0.501 (0.145)& 0.0458|0.366 & 0.514 (0.153)&0.579  & 0.541 (0.178)& 0.842 \\
IN   & 23.000 (5.587)& 1.762|14.097    &  28.540 (7.085)  & 26.815 & 51.860 (13.530)& 64.010\\
\hline
\end{tabular}
\end{center}
\caption{Integrated cross section of TCS in units of nb for $ep$ collisions for ${(M^2_{\ell^+\ell^-})}_{\mathrm{min}}= 0.1 \,\,(1.0)  \ GeV^2$. The event rates per year are also presented for the cut $M^2_{\ell^+\ell^-}> 1$ GeV$^2$.}
\label{tab:2}
\end{table*}

As already mentioned, the calculations for $ep$ collisions will be  performed at the center-of-mass energies of the proposed future facilities as the LHeC, its High-Luminosity (HL-LHeC) and High-Energy (HE-LHeC) updates and the Future Circular Collider in the lepton-hadron mode (FCC-eh). Their values of collision energies are outlined in Table \ref{tab:1} \cite{Bordry:2018gri}. In Fig. \ref{fig:dsdmep},  the differential cross sections of dilepton invariant mass, Eq. (\ref{eq:difXsec}), are presented for the design  energies of the projected experiments mentioned. Furthermore, we calculated the integrated cross section for the $\gamma p \to \ell^+ \ell^- p $ process over an interval between $(M^2_{\ell^+ \ell^-})^{min}$ and infinity. The values as a function of minimum dilepton invariant mass,  $(M^2_{\ell^+ \ell^-})^{min}= 0.1 \ GeV^2$ and $(M^2_{\ell^+ \ell^-})^{min}= 1 \ GeV^2$ are summarized in Table \ref{tab:2}, respectively. We also provide the event rates per year for $M^2_{\ell^+ \ell^-} > 1 \ GeV^2$ using the design luminosities at each energy \cite{Bordry:2018gri}.  On average, the behavior for MPM, GBW and KS are quite similar in the dilepton mass interval considered. The IN model produces a higher cross section with sizable deviation compared to the other UGDs. This fact and the matter of GBW prediction being very close to the remaining UGDs mean that this process is dominated by transverse momentum  around critical line, $Q_s$, or so. We call attention once again for the difference on the transition hard-soft in the IN model, which occurs at a fixed scale having order of magnitude of a few GeV. At very high energies, the saturation scale is enhanced a lot, $Q_s(W)\sim (W/W_0)^{0.12}$ GeV, and therefore in the saturation models (or saturation inspired parametrizations) there is an increasing contribution from transverse momenta in the region $k_T\lesssim Q_s(W)$. 

The presented cross section can be directly compared to previous calculations in literature using the color dipole formalism or $k_T$-factorization. Within the color dipole picture in the spacelike approximation \cite{Machado:2008zv,Mariotto:2013qsa}, it was found that the invariant mass distribution  is driven by the small mass region and the forward amplitude scales with $\sim Q_s^2(x)/M_{\ell^+\ell^-}^2$ when parton saturation models are considered with the critical line being defined by the saturation scale, $Q_s(x)$. This leads to a differential cross
section having the behavior $d^2\sigma /dM_{\ell^+\ell^-}^2 \sim \frac{1}{M_{\ell^+\ell^-}^6}
[1+\ln(M_{\ell^+\ell^-}^2)]$. The integrated cross section was predicted to be 0.08 nb using the cut $M_{\ell^+\ell^-} \geq  1.5$ GeV \cite{Mariotto:2013qsa} for $W_{\gamma p} = 1.4 $ TeV. Having in mind that the spacelike cross section is 3 or 4 times smaller than the timelike one, that calculation is similar to ours for the LHeC/HL-LHeC energy using the models based on saturation physics. Interestingly, the inclusive dilepton photoproduction  has been also estimated in Ref. \cite{Mariotto:2013qsa}, where it was found to be $\sigma_{inc}= 0.78$ nb for the integrated cross
section integrated for $M_{\ell^+\ell^-} \geq 3$ GeV. The first calculation using $k_T$-factorization was performed in Ref. \cite{Schafer:2010ud} using only the IN distribution. The present calculation is fully consistent with that study, with the integrated cross section for HL/HE-LHeC energies being $\sim 0.5$ nb for $(M^2_{\ell^+ \ell^-})_{\mathrm{min}} \geq 1$ GeV.

Moving now to $pp$ collisions, Fig. \ref{fig:dsdmpp} presents the dilepton invariant mass distribution  from TCS process at the LHC (Large Hadron Collider) energies of 7 TeV (left panel) and 13 TeV (right panel) at central rapidity (y=0). In addition, Table \ref{tab:3} shows the cross sections for the LHC13, HE-LHC (27 TeV) and FCC-hh (100 TeV) \cite{Bordry:2018gri}. As the predictions for the HL-LHC (14 TeV) are quite similar to the LHC13 ones, we are only displaying the latter. As expected, the behavior follows the pattern verified in $ep$ collisions. Once again, the differential cross section is dominated by dileptons produced with low invariant mass and there is a large dispersion in the absolute value of the cross section when different models are considered.  The results can be contrasted with the phenomenological predictions for dilepton production coming from the two-photon channel and hard diffractive channel given by Pomeron-Pomeron interactions. The invariant mass distribution has been predicted for 14 TeV in \cite{Kubasiak:2011xs}, where it was found to be $d\sigma/dM_{\ell^+ \ell^-} \simeq 0.7$ nb for $M_{\ell^+ \ell^-}=2 $ GeV and $d\sigma/dM_{\ell^+ \ell^-} \simeq 0.1$ pb for $M_{\ell^+ \ell^-}=10 $ GeV considering the IN UGD. Our results are similar at 13 TeV, where the evaluations using models based on saturation physics give systematically smaller values than the IN parametrization. The same is true for the rapidity distribution, as we will see in what follows. Still, the invariant mass distribution from the two-photon production is around two orders of magnitude higher than the exclusive production through the photon-Pomeron channel \cite{Kubasiak:2011xs}.

\begin{table}[t]
\begin{center}
\begin{tabular}{|l|c|c|c|}
\hline
 & LHC13 ($\times 10^{-2}$) & HE-LHC ($\times 10^{-2}$) & FCC-hh ($\times 10^{-2}$) \\
\hline
GBW    & 1.267 & 1.599 & 2.365     \\
MPM    & 2.272 & 3.133	& 4.024   \\
KS     & 0.640 & 0.790 & 1.085 \\
IN     & 6.653 & 9.530 & 18.410 \\
\hline
\end{tabular}
\end{center}
\caption{Cross section $d\sigma /dy$ of TCS in units of nb for $pp$ collisions at central rapidity (y=0) integrated for ${(M^2_{\ell^+\ell^-})}> 1.0  \ GeV^2$.}
\label{tab:3}
\end{table}

\begin{figure}[t]
\centering
    \includegraphics[width=1.0\linewidth]{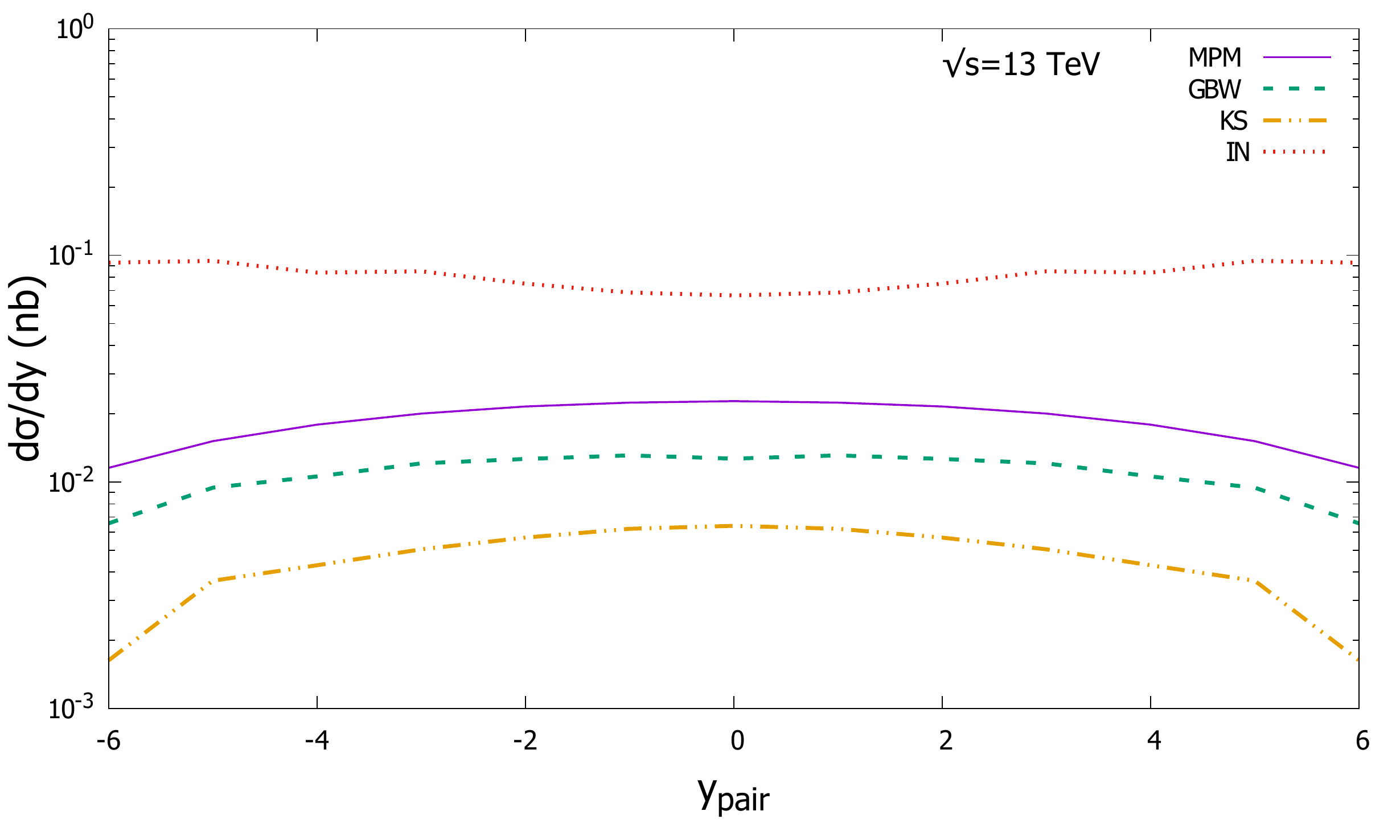}
    \caption{Dilepton rapidity distribution of TCS cross section for $pp$ collisions at the LHC for  $\sqrt{s} = 13$ TeV.}
    \label{fig:dsdy}
\end{figure}

Finally, in Fig. \ref{fig:dsdy}  the  rapidity distribution for $pp$ collisions at the LHC is displayed for the energy of 13 TeV.  At midrapidity, the cross section ranges in $d\sigma/ dy (y=0) \simeq 10 - 100$ pb and contains huge theoretical uncertainty. This is due to the amplification of deviations coming from different model assumptions for a cross section for an exclusive observable. Namely, the large uncertainty can be traced back to the values of cross sections evaluated in quantities proportional to $\phi (x,k)$ squared. In Ref. \cite{Kubasiak:2011xs}, where only the IN UDG has been utilized, the prediction at 14 TeV is 30 pb for midrapidity. This value is consistent with our calculations in its order of magnitude. The authors in \cite{Kubasiak:2011xs} found that the contribution from the process $\gamma \gamma \rightarrow  \ell^+\ell^-$ at central rapidity is around 10 nb, whereas the  contribution from inclusive single diffractive and central diffractive production of dileptons reaches 1 nb. Similar predictions are also presented in Ref. \cite{Goncalves:2018gca} at 13 TeV, concerning the two-photon and inclusive diffraction channel. In that study, the Forward Physics Monte Carlo (FPMC) has been used and the role played by pair transverse momentum cuts was demonstrated in order to disentangle the exclusive photon-induced production at low-$p_T$ from the diffractive sector.  Our conclusions about the exclusive dilepton production in $pp$ collisions are the same of those presented in \cite{Kubasiak:2011xs}, where the cross section for exclusive diffractive production is almost the order of magnitude  than that for the central diffractive production mechanism. A comprehensive analysis is needed (combining $p_T$ and $M_{\ell^+\ell^-}$ cuts) in order to demonstrate the feasibility of a experimental measurement.

The same process in $pp$ ultraperipheral collisions has been also investigated within the GPD formalism. The prediction for the integrated cross section using NLO GRVGJR2008 PDFs at hard scale $\mu_F^2 = 5 $ GeV$^2$ is 1.9 pb at 14 TeV \cite{Pire:2008ea}, which has the order of magnitude similar to our predictions for the saturation models like the KS UGD. However, the general trend is the estimates using $k_T$-factorization being higher than  those from GPD formalism. The background from the Bethe-Heitler process was estimated to be 2.9 pb \cite{Pire:2008ea}, which cannot be neglected even in high energy scattering. From the experimental point of view, the dilepton production coming from QED contribution is well understood. For instance, the ATLAS collaboration has measured \cite{Aad:2015bwa} the cross section at 7 TeV in the electron channel, which is determined to be $\sigma (\gamma\gamma \rightarrow e^+e^-) =  0.428 \pm 0.039$ pb, whereas in the muon channel one has $\sigma (\gamma\gamma \rightarrow \mu^+\mu^-) = 0.628 \pm 0.038$ pb (errors summed into quadrature, $p_T\,\gsim\, 10$ GeV and $|\eta|<2.4$). The LHCb collaboration has measured in the dimuon channel the cross section $\sigma (pp\rightarrow \mu^+\mu^-pp) = 67\pm 19 $ nb  \cite{LHCb-CONF-2011-022} (errors summed into quadrature, $M>2.5$ GeV and $\eta_{\mu^-}<2.4$).

\section{Summary and Conclusions}

In this work we calculated the dilepton invariant mass distribution from the TCS cross section in $ep$ collisions for the LHeC, HL-LHeC, HE-LHeC and FCC-eh energies.  Besides, we also evaluated this observable for $pp$ collisions at the LHC7, LHC13, HL-LHC, HE-LHC and FCC-hh, along with the rapidity distribution in the case of the LHC13. It was found that the theoretical uncertainty is quite large when we consider different models for the UGDs including those containing parton saturation effects. We found a deviation around one order of magnitude in the models considered in present study. There are other uncertainties coming from the $t$ behavior of the non-forward amplitude, the ansatz for the skewedness corrections and the prescription for the coupling at very low dipole invariant mass. It is clear that a comprehensive analysis on the kinematics variables should be performed in order to disentangle experimentally the TCS contribution from the similar signal coming from Bethe-Heitler background and also from exclusive dilepton production in two-photon fusion. The exclusive diffraction processes are currently being investigated by CMS and ATLAS collaborations at the LHC \cite{Royon:2020uoo} and the study presented here is complementary to the usual predictions in two-photon physics and inclusive diffraction (see a review in \cite{Royon:2020soh}). This interesting subject definitely needs more work and the analysis for nuclear targets is ongoing and will be presented in future contribution.
\label{summary}

\section*{Acknowledgments}

This work was financed by the Brazilian funding
agencies CNPq and  CAPES. We appreciate the kind support of Hannes Jung for helping us to install and use the  TMDlib. We also thank  Igor Ivanov for providing us his unintegrated gluon distribution code. We are grateful to  Wolfgang Sch\"afer for useful discussions and correspondence.

\bibliographystyle{h-physrev}
\bibliography{referencias_tcs}

\end{document}